# APPLICATION OF ARTIFICIAL NEURAL NETWORK IN MARKET SEGMENTATION: A REVIEW ON RECENT TRENDS

Manojit Chattopadhyay[1]*, Pranab K Dan[2], Sitanath Majumdar[3], Partha Sarathi Chakraborty[4]

[1] Dept. of Computer Application, Pailan College of Management & Technology, Kolkata-700104

[2] School of Engineering, West Bengal University of Technology, Kolkata-700064

[3] Dept. of Business Management, University of Calcutta, Kolkata-700027

[4] Dept. of Business Management, Pailan College of Management & Technology, Kolkata-700104

[1]* Corresponding author:  mjc02@rediffmail.com

## Abstract

Despite the significance of Artificial Neural Network (ANN) algorithm to market segmentation, there is a need of a comprehensive literature review and a classification system for it towards identification of future trend of market segmentation research. The present work is the first identifiable academic literature review of the application of neural network based techniques to segmentation.

Our study has provided an academic database of literature between the periods of 2000–2010 and proposed a classification scheme for the articles. One thousands (1000) articles have been identified, and around 100 relevant selected articles have been subsequently reviewed and classified based on the major focus of each paper.

Findings of this study indicated that the research area of ANN based applications are receiving most research attention and self organizing map based applications are second in position to be used in segmentation. The commonly used models for market segmentation are data mining, intelligent system etc.

Our analysis furnishes a roadmap to guide future research and aid knowledge accretion and establishment pertaining to the application of ANN based techniques in market segmentation. Thus the present work will significantly contribute to both the industry and academic research in business and marketing as a sustainable valuable knowledge source of market segmentation with the future trend of ANN application in segmentation.

# 1.    INTRODUCTION

Market segmentation has become a fundamental concept both in marketing theory and in practice (Smith, 1956). Wendell (1956), an American marketing researcher first presented the concept of market segmentation and the concept was further developed by many ensuing scholars and adopted by many companies. Market segmentation can be described as the process of partitioning a large market into the smaller groups or the clusters of customers (Weinstein, 1987; Smith, 1956; Kotler & Gordon, 1983; Croft, 1994; Myers, 1996). The similarity of purchase behavior can be identified from the similarities within each segment.

Market segmentation has several benefits over mass marketing as it provides the opportunity to expand a market after satisfying the specific needs of particular consumers (MacQueen et al., 1967).The information of the segments helps the decision makers to reach all customers effectively with one basic marketing mix (Anderson & Vincze, 2000). Thus it helps them to deploy resources more effectively and efficiently and determine the particular competitive strategies (i.e. differentiation, low cost, or focus strategy) (Aaker, 2001). Segmentation helps to build closer relationships between the customers and the company. Where the company can create a more fine-tuned product or service offering and price it suitably for the target segment (Kotler, 2000) for which the company can more easily select the best distribution and communications channels, and also have a clearer picture of its competitor.

The strong conceptual and theoretical works paved the path for the subsequent empirical work through the integration and extension of the review of the past empirical work (Stewart and

Zinkhan, 2006). Thus, the present work presents an analysis with future trend of Artificial Neural Network (ANN) techniques related to market segmentation after a comprehensive review of literature published in academic journals between 2000 and 2010. A classification of framework is also presented. The paper is organized as follows: first, the research methodology used in the study is described; second, the method for classifying ANN articles in segmentation is presented; third, articles about ANN in segmentation are analyzed and the results of the classification are reported; and finally, the conclusions, limitations and implications of the study are discussed.

## 2. RESEARCH METHODOLOGY

As the nature of research in ANN techniques in segmentation are difficult to confine to specific areas, the relevant materials are scattered across various journals. Expert systems (ES) with application and Information system are the most common discipline for ANN research in market segmentation. Consequently, the following online journal databases were searched to provide a comprehensive bibliography of the academic literature on ANN research in market segmentation:

- Google Scholar

- Citeceerx

- Social Science Network

- DOAJ

- Emerald Fulltext

- Ingenta Journals

- Science Direct

- IEEE Transaction.

The literature search was based on the query strings, ''market segmentation", "target marketing" and ''artificial neural network", which originally produced approximately 900 articles. The full text of each article was reviewed to eliminate those that were not actually related to application of ANN techniques in segmentation. The selection criteria were as follows:

- Only those articles that had been published in Neural Network, Expert System, Information system, decision support system, and data mining or marketing related journals were selected, as these were the most appropriate outlets for ANN techniques in segmentation and the focus of this review.

- Only those articles which clearly described how the mentioned ANN technique(s) could be applied and assisted in market segmentation were selected.

- Conference papers, masters and doctoral dissertations, textbooks and unpublished working papers were excluded, as most often researchers use journals to gain information and publish new findings. Nord & Nord (1995) reported journals represent the highest level of research.

--------------------------Figure 1----------------------------- Figure 2---------------------------------

## 3. CLASSIFICATION FRAMEWORK

For most business firms, locating and specifically targeting unique market segments is both a reality and a necessity in today's competitive marketplace. In this new and competitive commercial framework, market segmentation techniques can give marketing researchers a leading edge: because the identification of such segments can be the basis for effective targeting and predicting of potential customers (O'Connor and O'Keefe, 1997). According to Wedel and Kamakura (2000), market segmentation methods can be largely classified based on two criteria for the four categories: a priori or post hoc, and descriptive or predictive statistical methods. When the type and number of segments are determined in advance by the researcher then the a priori approach is used and when the type and number of segments are determined based on the results of data analyses then the post hoc approach is used. The post-hoc methods are relatively powerful and frequently used in practice (Dillon et al., 1993; Wedel and Kamakura, 1998). A single set of segmentation bases that has no distinction between dependent and independent variables are related with the descriptive methods. When one set consists of dependent variables to be explained or predicted by a set of independent variables then the predictive methods are applied.

The bases available for segmenting a market are nearly unlimited and can include product class behaviors, product class preferences, product class-related attitudes, brand selection behavior, brand-related attitudes, purchasers' attitudes toward themselves and their environment, demographics, geographics, and socioeconomic status. In general, there are four major classes of traditional algorithms for conducting traditional post hoc segmentation studies: Cluster analysis, Correspondence analysis, Search procedures, and Q-type factor analysis. Among clustering methods, the K-means method is the most frequently used (Anil et al., 1997).

Starting in the early 1990s, ANNs have been developed to address host of analytical problems. In general, ANNs are given a set of input variables and a set of known outcomes, and the algorithm is asked to find the best relationship between the inputs and the outputs. This is dome by initially on a subset of the data, called the learning set.  Then the algorithm uses one or more "hidden layers" between input nodes, or neurons, and adjusts the weight of each input to that neuron and accurately predicts the outcome. The outputs are validated with a third sample, the validation sample. An unsupervised neural network of the artificial neural networks (ANNs) where the outcomes are not a priori  have been recently applied to a wide variety of business areas (Vellido et al., 1999) including market segmentation (Balakrishnan et al., 1996; Kuo et al., 2002a, b). The Kohonan Self-Organizing Map of unsupervised ANN used in clustering for large and complex data. Self-Organizing Feature Maps (SOM), can project high dimensional input space on a low-dimensional topology, allowing one to visually determine out the number of clusters (Lee et al., 1977; Pykett, 1978).

## 3.1 Classification Framework -ANN Dimension

A careful revision has been done for each article and the source of ANN application in market segmentation has been categorized into 8 dimensions as shown in Fig. 1 and Fig. 2 and fourteen categories of ANN algorithms (figure 3). Although this search was not exhaustive, it serves as a comprehensive base for an understanding of ANN research in segmentation.

The application of neural networks in marketing area is relatively new but is becoming popular because of their ability of capturing nonlinear relationship between the variables. Numerous applications of NNs models in marketing discipline are available, to mention a few are Market Segmentation, Market Response Prediction, New Product Launch, Sales Forecasting, Consumer Choice Prediction, etc. Market modeling is an extremely important issue in marketing. At the aggregate level, market share models are commonly used in marketing for a number of different purposes. These include the estimation of price and advertising elasticities as well as more generally, predicting the effects of changes in marketing variables. Cluster analysis is a common tool for market segmentation. Conventional research usually employs the multivariate analysis procedures. Kuo et al., (2002a,b, 2006) compared three clustering methods and proposed that SOM performs better clustering than the other conventional methods. Lewis et al., (2001) successfully implemented SOM based neural network to Identify the residential property sub-markets. A data mining associatiation ruile based on SOM has been developed and applied to a sample of sales records from database for market fragmentation (Changchien and Lu, 2001). Mazanec(2001) applied a topology sensitive vector quantization method for market structure analysis. Hopefield neural network found to be more useful to retailers for segmenting markets(Boone and Roehm, 2002). Bayesian neural networks offer a viable alternative for purchase incidence modeling to identify whether a customer uses the credit facilities of the direct mailing company (Baesens et al., 2002). Adaptive resonance theory, particularly ART2, neural network has been applied as a toolkit to develop a strategy for acquiring customer requirement patterns (Chen et al., 2002). Kauko et al.,(2002) researched on neural network modelling using SOM and Learning Vector Quantization(LVQ) with an application to the housing market of Helsinki, Finland which shows how it is possible to identify various dimensions of housing submarket formation by uncovering patterns in the dataset, and also shows the classification abilities of two neural network techniques. A genetic algorithm approach has been adopted to ensure that customers in the same cluster have the closest purchase patterns (Tsai and Chiu, 2004). A two stage method comprising SOM and GA (genetic algorithm) deployed for wireless telecommunications industry market segmentation and the results showed that the proposed method has better output in segmentation (Kuo et al., 2004). A GA based intelligent system approach found to be efficiently used in customer targeting (Kim and Street, 2004). The back propagation neural network based application deployed in Cape Metropolitan Tourism data has been shown to track the changing behavior of tourists within and between segments (Bloom, 2005). It was found that NN models outperforms the multinomial logut model in determining the most profitable time in a purchasing history to classify and target prospective consumers new to their categories (Kaefer et al., 2005). Kim et al., (2005) deployed an ANN guided by genetic algorithms (GAs) successfully to target households. Targeting of customer segments with tailored promotional activities is an important aspect of customer relationship management. Kiang et al.,(2006) applied the SOM networks to a consumer data set from American Telephone and Telegraph Company (AT&T) and the research established that the SOM network performs better than the two-step procedure that combines factor analysis and K-means cluster analysis in uncovering market segments. Reutterer et al., (2006) empirically established evidence of significant positive impacts on both profitability and sales for segment-specific tailored direct marketing campaigns. Huang et al., (2007) implemented support vector clustering (SVC), SOM and k-means in a case study of a drink company and on the basis of the numerical results; they found that SVC outperforms the other methods. SOM has been shown to generate significant segments with particular characteristics are found in the mature market (Bigné et al., 2008). A smaller market can also be targeted using the Marketing segmentation which is useful for decision makers to reach all customers effectively with one basic marketing mix. A backpropagation neural network model has been shown to be useful in identifying existing patterns of hospitals' consumers (Lee et al., 2008). Chan (2008) demonstrated that the segmentation using GA based method can more effectively target valuable customers than random selection. ANN and particle swarm optimization methods has been implemented to generate precise market segmentation for marketing strategy decision making and extended application (Chiu et al., 2009). Wang (2009) adopted a hybrid kernel-based clustering techniques for outlier identification and robust segmentation in real application. Hung and Tsai (2008) successfully implemented a novel market segmentation approach, using the hierarchical self-organizing segmentation model (HSOS), for dealing with a real-world data set for market segmentation of multimedia on demand in Taiwan. Fuzzy Delphi method, self-organizing maps (SOM) and a visualization techniques have been developed and successfully implemented to cluster customers according to their various characteristic variables and visualize segments by producing colorful market maps (Hanafizadeh and Mirzazadeh,2011). A firm that operates in multiple regions, a market segmentation method that can integrate data from different regions to obtain a set of generalized segmentation rules can greatly influence the competitiveness of the company. Mo et al., (2010) applied self-organizing map (SOM) network technique as both a dimension reduction and a clustering tool to market segmentation using the data from one of the largest credit card issuing banks in China that includes surveys of customer satisfaction attributes and credit card transaction history and have shown that SOM for multi-region segmentation is an effective and efficient method compared to other approaches. An expert system has been developed by the researchers using a Fuzzy Delphi method and a back propagation neural network model and has been applied to data of customers for perfume selection with more correct classification rate (Hanafizadeh et al., 2010). Chen et al., (2010) applied a 3NN+1 based CBR system by means of genetic algorithms to a real case of notebook market to demonstrate its usefulness for market segmentation. From the results of the real case, it has been shown that the system would be valuable to enterprises, for developing marketing strategies.

3.2 Classification Process

All the selected articles were individually reviewed and categorized based on the proposed classification framework by the authors of this paper. The proposed classification scheme is consisting of the following phases:

- Online database search

- Initial classification by the researcher

- Verification of the classification result

If there is any discrepency found in the above steps then a unanimous decision has been taken in this regard to include in the final classification scheme. The procedure of the proposed scheme of classification is shown in figure 3. The collection of articles was analysed on the basis of ANN application in market segmentation, by year of publication, author and according to journal in which article was published. After Critically reviewing the literature on ANN application applied to market segmentation a classification framework a) on domains of the journals, b) year of publication and c) articles published in each individual journal along with d) a trend of ANN techniques applied in segmentation process have been identified.

---------------------------- ------------------Figure 3 here----------------------------------------------

## 4. CLASSIFICATION OF THE ARTICLES

A detailed distribution of the 64 articles categorized by the proposed classification scheme is shown in the Table 1 based on year of publication, journal name, title, ANN tools used, and authors. The following 14 types of ANN algorithms (Table 1) are found to be applied on market segmentation research from the year 2000 to 2010 in the selected reviewed journals: i) NN algorithm, ii) Meta Heuristic tools, iii) ARNN(association reasoning neural network), iv) ART2

(adaptive resonance theory), v) Bayesian NN, vi) Back Propagation NN, vii) Data Mining, viii) hybrid fuzzy tools, ix) Genetic Algorithm(GA), x) hopefield NN, xi) hybrid NN, xii) Self Organizing Map(SOM), xiii) support vector machine (SVM), xiv) Vector Quantization.

-----------------------------------------Table 1 -----------------------------------------------

The figure 4 shows the distribution of articles published in each year. The year 2004 and 2010 shows the highest number articles (10) and next highest 9, 8 in the year 2008 and 2002 respectively published on ANN application to market segmentation from the selected 64 articles. There is an increasing trend of articles published in the last three years 2008 to 2010 (total 26 publications).

-----------------------------------------------Figure 4------------------------------------------------------

## 4.1. Distribution of Articles by ANN Models

The figure 5 shows the identified 14 category of ANN application to market segmentation. The distribution of articles published in each ANN category also revealed that self organizing map (22) has been applied in most of the selected 64 articles i.e. 34.38% of the total articles published. NN algorithm (11) and GA (8) is the next highest ANN tool applied in segmentation research contributing 17.19% and 12.50% of the total articles with ANN tool found respectively.

-------------------------------------------------Table 2-------------------------------------------------

## 4.2. Distribution of Articles by Year of Publication

Table 2 shows the distribution of articles published with 14 categories of ANN techniques applied in each year. It also revealed that most of the articles (10) with ANN tools published in the year 2004 and 2010 (15.63%). In the year 2008 and 2002 total 9 (14.06%) and 8 (12.50% ) articles published respectively. Highest number of articles published with the application of SOM (22) and NN (14). The distribution of articles with ANN application by year of publication is shown in Fig. 6. It is obvious that publications which are related to application of ANN techniques in market segmentation have increased significantly from 2008 to 2010. In 2003, 2006, 2005 and 2009 the amount of publication decreased by 87.50%, 50%, 40% and 22.22% respectively when compared to their previous year. There is a highest increase in publication in the year 2004 compared to the year 2003 (900%) and next highest increase are 125%, 42.86% in the year 2008 and 2010 respectively compared to their previous year.

----------------------Figure 5------------------------------- Figure 6--------------------------------------

### 4.3. Distribution of Articles by Journal in Which the Articles were Published

The figure 7 shows the distribution of 28 journals selected after review with the number of articles related to ANN application in segmentation published in each journal. The figure 7 also shows that the journal Expert system with Applications published highest number of articles (29) related to the criteria of our research..

--------------------------------------------Figure 7 here------------------------------------------------

The Table 3 shows the distribution of articles of ANN algorithm applied to segmentation by Journals. From the table the highest number of articles published is shown in the journal Expert Systems with Applications contributing 45.31% of the total articles selected after review.

--------------------------------------------------Table 3 here----------------------------------------------------

### 5. CONCLUSION, RESEARCH IMPLICATIONS AND LIMITATIONS

Application of artificial neural network techniques in market segmentation is an emerging inclination in the industry and academia. It has paying the attention of researchers, industry practitioners and academics. This work has identified sixty four articles related to application of artificial neural network techniques in market segmentation, and published between 2000 and 2010. This article aims to give a research review on the application of neural network in the market segmentation domain and techniques which are most often used. While this review work cannot claim to be exhaustive, but it presents reasonable insights and shows the prevalence of research on this area under discussion. The outcomes obtainable from this article have a number of significant implications:

- Research on the application of neural network in segmentation will increase significantly in the future based on past publication rates and the increasing interest in the area.

- The majority of the reviewed articles relate to self organizing map i.e., 34.38% (22 articles) and 21.88% (14 articles) are related common neural network algorithms.

- The journal Expert Systems with Applications related to expert and intelligent systems in industry, government and university worldwide, contains 29 articles contributing 45.31% of the total 64 selected articles. Rest of the journals published more or less same number of articles with the count of 1 to 3. Thus a trend of ANN research to segmentation is more obvious from the articles published in the kind of journal related to expert system development.

- These articles could provide insight to organization strategists on the familiar artificial neural network practices used in market segmentation. Of the 64 articles related to 29 journals, highest number of journals (8) is in the domain related to management/marketing and information technology or computer science respectively. But the number of articles published is highest in the later domain of the information/expert system (60%).

- There are relatively fewer articles with the metaheuristic, ART2, data mining, Genetic Algorithm and fuzzy algorithms. Despite the fewer number of articles related to the above category of artificial neural network application to market segmentation, it does not mean the application of artificial neural network in this aspect is less mature than in the others. Applications of those algorithms in other domains, such as clustering or classification, may also be applied in segmentation if they possess the same purpose of analysing the distinctiveness of customers/market.

- The k-means clustering model is the most commonly applied model in segmentation by partitioning a large market into the smaller groups or the clusters of customers. This is not surprising as classification modeling could be used to find out the similarity of each segment to predict the effectiveness of similar purchase behavior of customers.

- With respect to the research findings, we suggest more research can be conducted in the market segmentation domain. In order to maximize an organization's profits through segmentation, strategists have to both segment the market and thus increase the profitability of the organisation.

This review work might have some limitations. Firstly, this work only surveyed articles published between 2000 and 2010, which were obtained based on a keyword query of ''market segmentation'' and ''artificial neural network''. Research papers could not be extracted which mentioned the application of artificial neural network techniques in segmentation but without a

keyword index. Secondly, this work restricted the query search for articles to 7 online repositories. The presence of other academic journals may be able to provide a more comprehensive representation of the articles related to the application of artificial neural network in segmentation. Lastly, non-English publications were not considered in this study. We believe research regarding the application of artificial neural network techniques in segmentation have also been discussed and published in other languages.

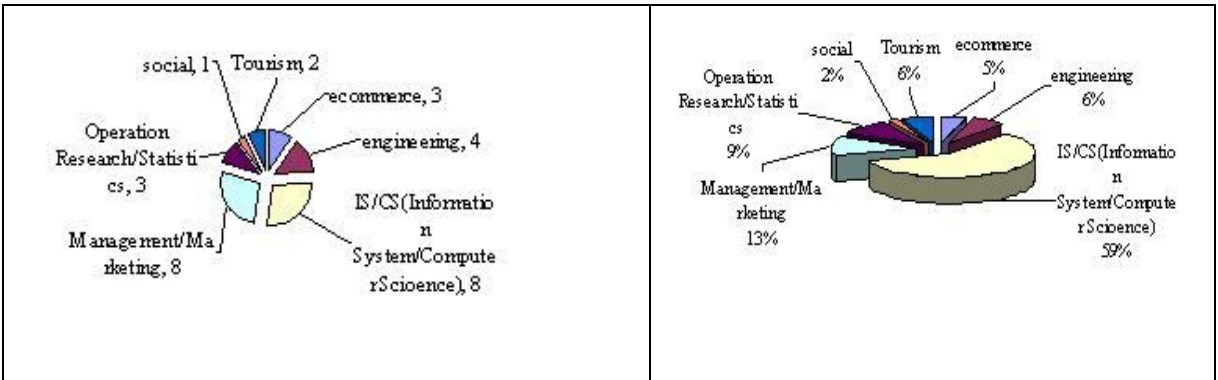

Figure 1. Number of Journals and their identified domains     Figure 2. % of articles published in each domains

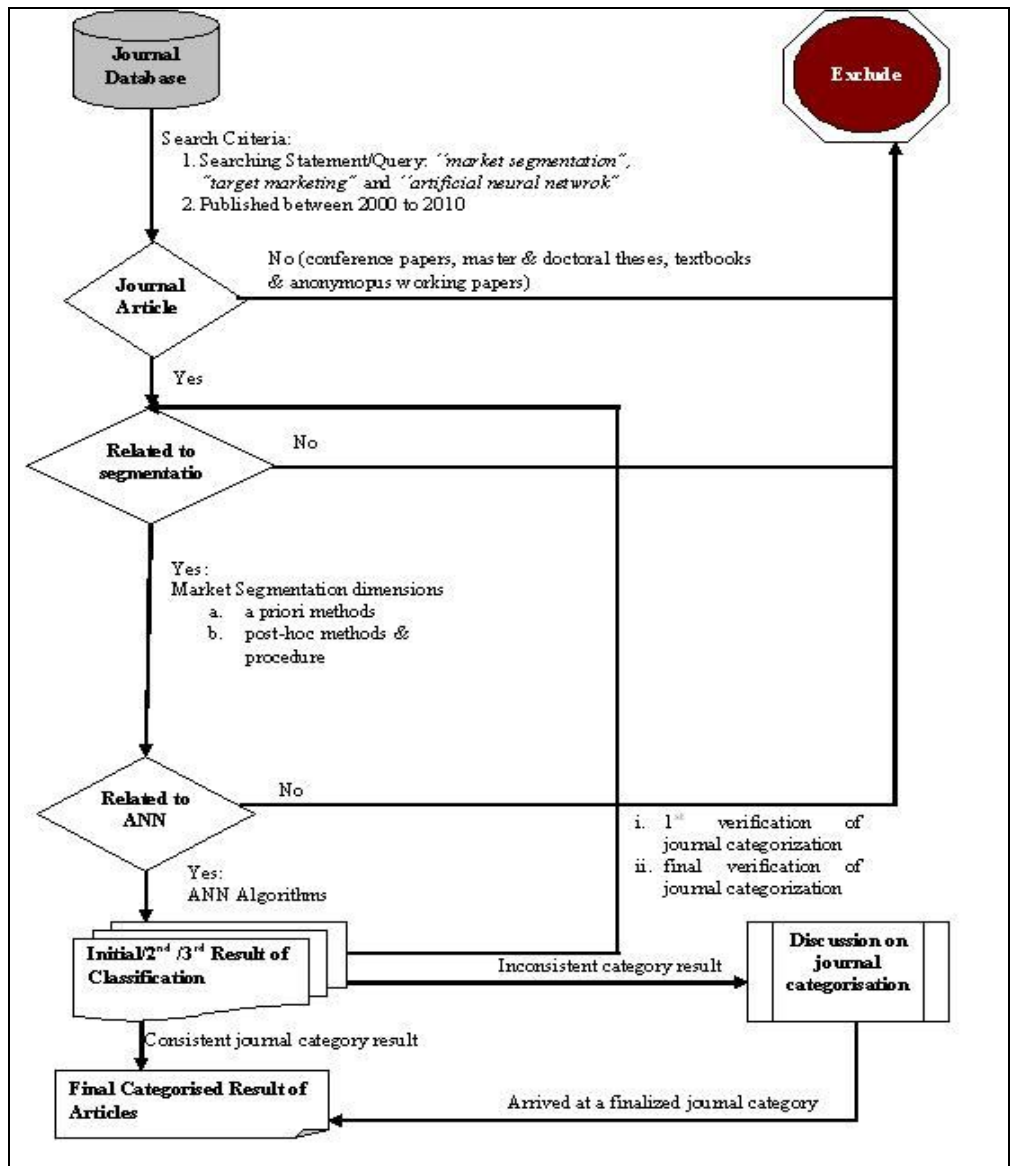

Figure 3. Proposed classification process

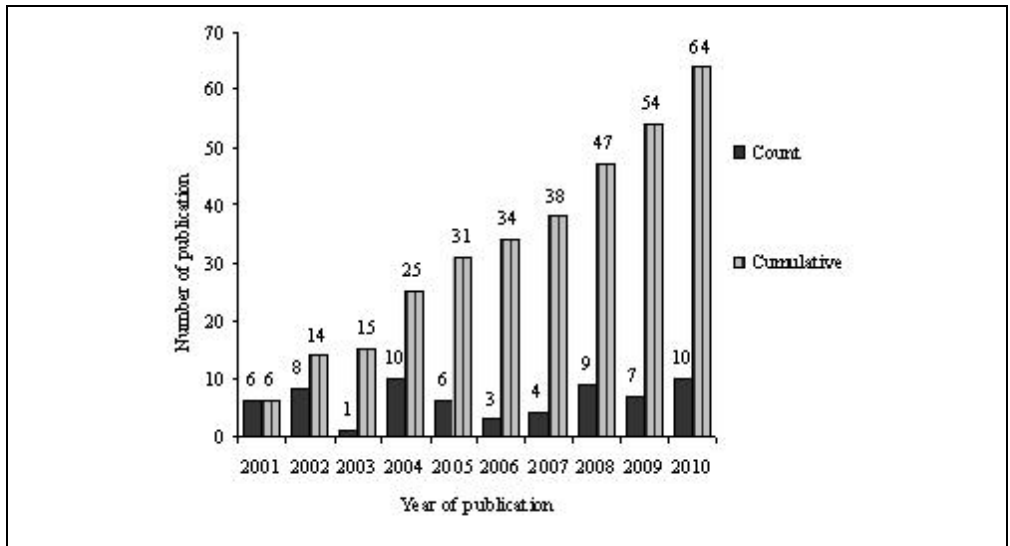

Figure 4. Number of articles published in each year along with cumulative count of articles

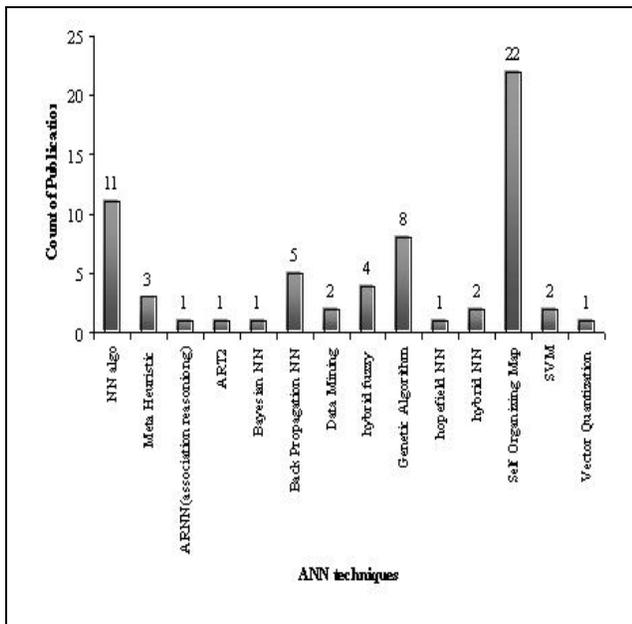

Figure 5 Articles published in each category of ANN-based algorithms applied to Market Segmentation

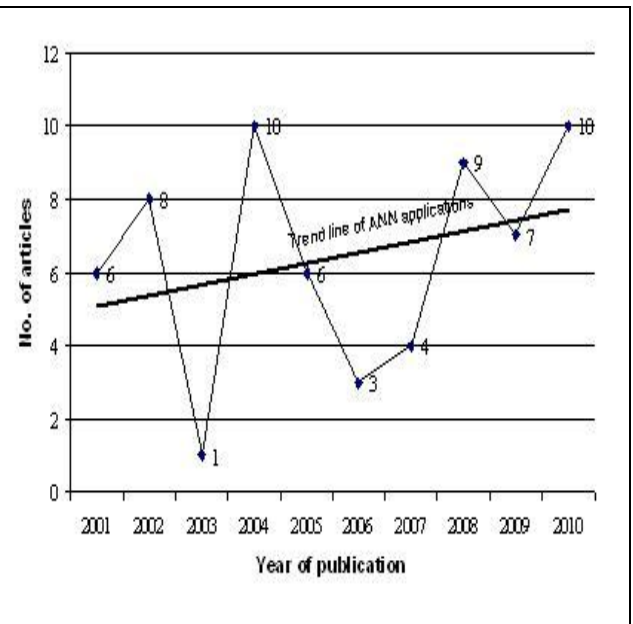

Figure 6. Distribution of articles by year of publication

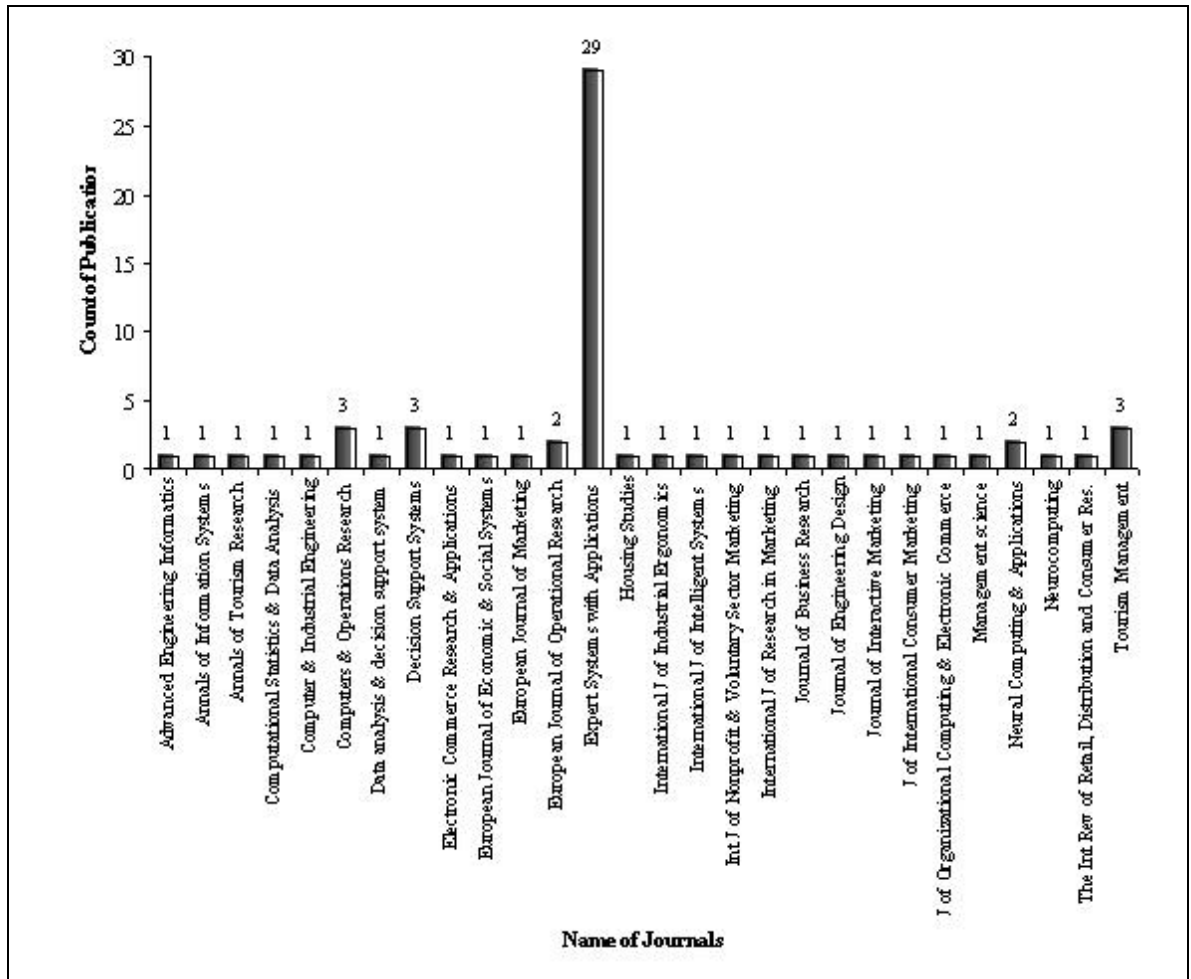

Figure 7. Articles published on ANN application in segmentation in 29 reviewed Journals

Table 1 Distribution of articles by journal name, title, ANN tools used, author and year of publication

| Year of Publication | Journal Name | Title | ANN Used | Tools | Authors |
|---|---|---|---|---|---|
| 2001 | European Journal of Economic and Social Systems | Buying behavior study with basket analysis: pre-clustering with a Kohonen map | SOM | | Pierre Desmet |
| 2001 | European Journal of Marketing | Neural market structure analysis: Novel topology-sensitive methodology | VQ | | Josef A. Mazanec |
| 2001 | Expert Systems with Applications | Mining association rules procedure to support on-line recommendation by customers and products fragmentation | SOM | | S. Wesley Changchien and Tzu-Chuen Lu |
| 2001 | International Journal of Intelligent Systems | Knowledge discovery in a direct marketing case using least squares support vector machines | SVM | | S. Viaene, B. Baesens, T. Van Gestel, J. A. K. Suykens, D. Van den Poel, J. Vanthienen, B. De Moor, G. Dedene |
| 2001 | Journal of International Consumer Marketing | Micro-Market Segmentation Using a Neural Network Model Approach | BPNN | | Jafar Ali, C. P. Rao |

| | | | | |
|---|---|---|---|---|
| 2001 | Neural Computing & Applications | Identification of Residential Property Sub-Markets using Evolutionary and Neural Computing Techniques | SOM | O.M. Lewis, J.A. Ware and D.H. Jenkins |
| 2002 | Advanced Engineering Informatics, | A strategy for acquiring customer requirement patterns using laddering technique and ART2 neural network | ART2 | Chun-Hsien Chen, Li Pheng Khoo, Wei Yan |
| 2002 | Computer & Industrial Engineering | Cluster analysis in industrial market segmentation through artificial neural network | SOM & K-means | R.J.Kuo, L.M.Ho,C.M.Hu |
| 2002 | Computers & Operations Research | Integration of self-organizing feature map and K-means algorithm for market segmentation | SOM | R. J. Kuo, L. M. Ho, C. M. Hu |
| 2002 | European Journal of Operational Research | Bayesian neural network learning for repeat purchase modelling in direct marketing | BNN | Bart Baesens, Stijn Viaene, Dirk Van den Poel, Jan Vanthienen, Guido Dedene |
| 2002 | Housing Studies | Capturing Housing Market Segmentation: An Alternative Approach based on Neural Network Modelling | NN | Tom Kauko, Pieter Hooimeijer, Jacco Hakfoort |
| 2002 | International Journal of Research in Marketing | Retail segmentation using artificial neural networks | HopefieldNN | Derrick S. Boone, Michelle Roehm |
| 2002 | Journal of Interactive Marketing | Retail segmentation using artificial neural networks | NN | DS Boone, M Roehm |
| 2002 | Tourism Management | Determinants of guest loyalty to international tourist hotels—a neural network approach | ANN | Sheng-Hshiung Tsaura, Yi-Chang Chiu, and Chung-Huei Huang |
| 2003 | Tourism Management | Segmenting the market of West Australian senior tourists using an artificial neural network | NN | Jaesoo Kim, Sherrie Weia and Hein Ruys |
| 2004 | Decision Support Systems | An intelligent system for customer targeting: a data mining approach | GA | Yong Seog Kim and W. Nick Street |
| 2004 | Expert Systems with Applications | An integrated data mining and behavioral scoring model for analyzing bank customers | SOM | Nan-Chen Hsieh |
| 2004 | Expert Systems with Applications | Segmentation of stock trading customers according to potential value | fuzzy k-means | H.W. Shin, S.Y. Sohn |
| 2004 | Expert Systems with Applications | A cross-national market segmentation of online game industry using SOM | SOM | Sang Chul Lee, Yung Ho Suh, Jae Kyeong Kim, Kyoung Jun Lee |
| 2004 | Expert Systems with Applications | Joint optimization of customer segmentation and marketing policy to maximize long-term profitability | MH | Jedid-Jah Jonker, Nanda Piersma and Dirk Van den Poel |
| 2004 | Expert Systems with Applications | A purchase-based market segmentation methodology | GA | C.-Y. Tsai, C.-C. Chiu |
| 2004 | International Journal of Industrial Ergonomics | Predicting automobile seat comfort using a neural network | ANN | M. Kolich |
| 2004 | Journal of Business Research | Using an artificial neural network trained with a genetic algorithm to model brand share | GA | Kelly E. Fisha, John D. Johnson, Robert E. Dorsey and Jeffery G. Blodgett |
| 2004 | Journal of Organizational Computing and Electronic Commerce | Integration of Self-Organizing Feature Maps and Genetic-Algorithm-Based Clustering Method for Market Segmentation | SOM & GA | R. J. Kuo; K. Chang; S. Y. Chien |
| 2004 | Tourism Management | Tourist market segmentation with linear and non-linear techniques | SOM | Jonathan Z. Bloom |
| 2005 | Annals of Tourism Research | MARKET SEGMENTATION: A Neural Network Application | SOM | Jonathan Z. Bloom |
| 2005 | Computers & Operations Research | A neural network application to consumer classification to improve the timing of direct marketing activities | ANN | Frederick Kaefer, Carrie M. Heilman, and Samuel D. Ramenofsky |

| 2005 | Data analysis and decision support system | The Number of Clusters in Market Segmentation | SOM | Ralf Wagner, Sören W. Scholz and Reinhold Decker |
|------|------|------|------|------|
| 2005 | Expert Systems with Applications | Constrained optimization of data-mining problems to improve model performance: A direct-marketing application | DM | Anita Prinzie, Dirk Van den Poel |
| 2005 | Management science | Customer Targeting: A Neural Network Approach Guided by Genetic Algorithms | GA | W. Nick Street, Filippo Menczer, Gary J. Russell |
| 2005 | The International Review of Retail, Distribution and Consumer Research | Market Basket Analysis by Means of a Growing Neural Network | ANN | Prof. Dr Reinhold Decker |
| 2006 | Decision Support Systems | An extended self-organizing map network for market segmentation—a telecommunication example | SOM | Melody Y. Kiang, Michael Y. Hu, Dorothy M. Fisher |
| 2006 | Expert Systems with Applications | Integration of self-organizing feature maps neural network and genetic K-means algorithm for market segmentation | SOM | R.J. Kuo, Y.L. An, H.S. Wang, W.J. Chung |
| 2006 | Expert Systems with Applications | Integration of self-organizing feature maps neural network and genetic K-means algorithm for market segmentation | SOM | R.J. Kuo, Y.L. An, H.S. Wang and W.J. Chung |
| 2007 | Computational Statistics & Data Analysis | The effect sample size on the extended self-organizing map network—A market segmentation application | SOM | Melody Y. Kianga, Michael Y. Hub, Dorothy M. Fisher |
| 2007 | Computers & Operations Research | Neural networks in business: techniques and applications for the operations researcher | NN | KA Smith, JND Gupta |
| 2007 | Expert Systems with Applications | Marketing segmentation using support vector clustering | SVM | Jih-Jeng Huang, Gwo-Hshiung Tzeng, Chorng-Shyong Ong |
| 2007 | Journal of Engineering Design | Market segmentation for product family positioning based on fuzzy clustering | fuzzyclustering | Yiyang Zhang Jianxin (Roger) Jiao; Yongsheng Ma |
| 2008 | Decision Support Systems | Modeling consumer situational choice of long distance communication with neural | NN | Michael Y. Hu, Murali Shanker, G. Peter Zhang, Ming S. Hung |
| 2008 | Expert Systems with Applications | Market segmentation based on hierarchical self-organizing map for markets of multimedia on demand | SOM | Chihli Hung, Chih-Fong Tsai |
| 2008 | Expert Systems with Applications | A recommender system using GA K-means clustering in an online shopping market | GA & K-means | K Kim, H Ahn |
| 2008 | Expert Systems with Applications | Variable selection in clustering for marketing segmentation using genetic algorithms | GA | Hsiang-Hsi Liu, Chorng-Shyong Ong |
| 2008 | Expert Systems with Applications | The exploration of consumers' behavior in choosing hospital by the application of neural network | BPNN | Wan-I Lee, Bih-Yaw Shih, Yi-Shun Chung |
| 2008 | Expert Systems with Applications | Intelligent value-based customer segmentation method for campaign management: A case study of automobile retailer | GA | Chu Chai Henry Chan |
| 2008 | Expert Systems with Applications | Selecting the right MBA schools – An application of self-organizing map networks | SOM | Melody Y. Kiang and Dorothy M. Fisher |
| 2008 | Neural Computing & Applications | Mature market segmentation: a comparison of artificial neural networks and traditional methods | SOM | Enrique Bigné, Joaquin Aldas-Manzano, Inés Küster and Natalia Vila |
| 2008 | Neurocomputing | Temporal self-organizing maps for telecommunications market segmentation | SOM | Pierpaolo D'Ursoa, and Livia De Giovanni |

| Year | Journal | Title | Method | Authors |
|---|---|---|---|---|
| 2009 | European Journal of Operational Research | Quantitative models for direct marketing: A review from systems perspective | NN & stat | Indranil Bose, Xi Chen |
| 2009 | Expert Systems with Applications | Applying artificial immune system and ant algorithm in air-conditioner market segmentation | ant algo | Chui-Yu Chiu, I-Ting Kuo, Chia-Hao Lin |
| 2009 | Expert Systems with Applications | Application of data mining techniques in customer relationship management: A literature review and classification | DM | EWT Ngai, L Xiu, DCK Chau |
| 2009 | Expert Systems with Applications | An intelligent market segmentation system using k-means and particle swarm optimization | swarm & k-means | Chui-Yu Chiu, Yi-Feng Chen, I-Ting Kuo, He Chun Ku |
| 2009 | Expert Systems with Applications | Neural networks and statistical techniques: A review of applications | NN | Mukta Paliwal, Usha A. Kumar |
| 2009 | Expert Systems with Applications | Outlier identification and market segmentation using kernel-based clustering techniques | hybridNN | Chih-Hsuan Wang |
| 2009 | International Journal of Nonprofit and Voluntary Sector Marketing | Psychographic clustering of blood donors in Egypt using Kohonen's self-organizing maps | SOM | Mohamed M. Mostafa |
| 2010 | Annals of Information Systems | Predicting Customer Loyalty Labels in a Large Retail Database: A Case Study in Chile | MLP | Cristián J. Figueroa |
| 2010 | Electronic Commerce Research and Applications | Identifying influential reviewers for word-of-mouth marketing | NN | Yung-Ming Li, Chia-Hao Lin, Cheng-Yang Lai |
| 2010 | Expert Systems with Applications | Visualizing market segmentation using self-organizing maps and Fuzzy Delphi method – ADSL market of a telecommunication company | SOM & fuzzy | Payam Hanafizadeh, Meysam Mirzazadeh |
| 2010 | Expert Systems with Applications | Apply robust segmentation to the service industry using kernel induced fuzzy clustering techniques | Hybrid FUZZY | Chih-Hsuan Wang |
| 2010 | Expert Systems with Applications | A two-stage clustering approach for multi-region segmentation | SOM | Jiahui Mo, Melody Y. Kiang, Peng Zou, Yijun Li |
| 2010 | Expert Systems with Applications | An expert system for perfume selection using artificial neural network | BPNN | Payam Hanafizadeh, Ahad Zare Ravasan and Hesam Ramazanpour Khaki |
| 2010 | Expert Systems with Applications | An expert system for perfume selection using artificial neural network | fuzzy delphi & bpn | Payam Hanafizadeh, Ahad Zare Ravasan, Hesam Ramazanpour Khaki |
| 2010 | Expert Systems with Applications | Bayesian variable selection for binary response models and direct marketing forecasting | BayesianNN | Geng Cui, Man Leung Wong, Guichang Zhang |
| 2010 | Expert Systems with Applications | Cosmetics purchasing behavior – An analysis using association reasoning neural networks | ARNN | I-Cheng Yeh, Che-hui Lien, Tao-Ming Ting, Yi-Yun Wang, Chin-Ming Tu |
| 2010 | Expert Systems with Applications | Application of a 3NN+1 based CBR system to segmentation of the notebook computers market | GA | Yan-Kwang Chen, Cheng-Yi Wang, Yuan-Yao Feng |

Table 2. Publication count of different categories of ANN techniques applied in each year

| | BPNN | SOM | SVM | VQ | NN | hybrid NN | ART2 | Hopefield NN | MH | hybrid fuzzy | GA | DM | Bayesian NN | ARNN | Total (Year wise) |
|---|---|---|---|---|---|---|---|---|---|---|---|---|---|---|---|
| 2001 | 1 | 3 | | 1 | 1 | | | | | | | | | | 6 |
| 2002 | 1 | 2 | | | 3 | | 1 | 1 | | | | | | | 8 |
| 2003 | | | | | 1 | | | | | | | | | | 1 |
| 2004 | | 4 | | | 1 | | | | 1 | 1 | 3 | | | | 10 |
| 2005 | | 2 | | | 2 | | | | | | 1 | 1 | | | 6 |
| 2006 | | 3 | | | | | | | | | | | | | 3 |
| 2007 | | 1 | 1 | | 1 | | | | | 1 | | | | | 4 |
| 2008 | 1 | 4 | | | 1 | | | | | | 3 | | | | 9 |
| 2009 | | 1 | | | 1 | 2 | | | 2 | | | 1 | | | 7 |
| 2010 | 2 | 2 | | | 1 | | | | | 2 | 1 | | 1 | 1 | 10 |

Table 3 Distribution of articles by journal

| *Journal Name* | *% of Articles* |
|---|---|
| Advanced Engineering Informatics, | 1.56% |
| Annals of Information Systems | 1.56% |
| Annals of Tourism Research | 1.56% |
| Computational Statistics & Data Analysis | 1.56% |
| Computer & Industrial Engineering | 1.56% |
| Computers & Operations Research | 4.69% |
| Data analysis and decision support system | 1.56% |
| Decision Support Systems | 4.69% |
| Electronic Commerce Research and Applications | 1.56% |
| European Journal of Economic and Social Systems | 1.56% |
| European Journal of Marketing | 1.56% |
| European Journal of Operational Research | 3.13% |
| Expert Systems with Applications | 45.31% |
| Housing Studies | 1.56% |
| International Journal of Industrial Ergonomics | 1.56% |
| International Journal of Intelligent Systems | 1.56% |
| International Journal of Nonprofit and Voluntary Sector Marketing | 1.56% |
| International Journal of Research in Marketing | 1.56% |
| Journal of Business Research | 1.56% |
| Journal of Engineering Design | 1.56% |
| Journal of Interactive Marketing | 1.56% |
| Journal of International Consumer Marketing | 1.56% |
| Journal of Organizational Computing and Electronic Commerce | 1.56% |
| Management science | 1.56% |
| Neural Computing & Applications | 3.13% |
| Neurocomputing | 1.56% |
| The International Review of Retail, Distribution and Consumer Research | 1.56% |